\documentclass[aps,pra,superscriptaddress]{revtex4}
\usepackage{mathrsfs}
\usepackage{epstopdf}
\usepackage{amsfonts}
\usepackage{graphicx}
\usepackage{amssymb}
\usepackage{amsmath}
\usepackage[retainorgcmds]{IEEEtrantools}
\begin{document}
\def\ran{\rangle}
\def\lan{\langle}
\def\cl{\centerline}
\def\bd{\begin{description}}
\def\be{\begin{enumerate}}
\def\ben{\begin{equation}}
\def\benn{\begin{equation*}}
\def\een{\end{equation}}
\def\eenn{\end{equation*}}
\def\benr{\begin{eqnarray}}
\def\eenr{\end{eqnarray}}
\def\benrr{\begin{eqnarray*}}
\def\eenrr{\end{eqnarray*}}
\def\ed{\end{description}}
\def\ee{\end{enumerate}}
\def\al{\alpha}
\def\b{\beta}
\def\bR{\bar\R}
\def\bc{\begin{center}}
\def\ec{\end{center}}
\def\dg{\dagger}
\def\d{\dot}
\def\D{\Delta}
\def\del{\delta}
\def\ep{\epsilon}
\def\g{\gamma}
\def\G{\Gamma}
\def\h{\hat}
\def\iny{\infty}
\def\La{\Longrightarrow}
\def\la{\lambda}
\def\m{\mu}
\def\n{\nu}
\def\noi{\noindent}
\def\Om{\Omega}
\def\om{\omega}
\def\p{\psi}
\def\pr{\prime}
\def\r{\ref}
\def\R{{\bf R}}
\def\ra{\rightarrow}
\def\up{\uparrow}
\def\dn{\downarrow}
\def\lr{\leftrightarrow}
\def\s{\sum_{i=1}^n}
\def\si{\sigma}
\def\Si{\Sigma}
\def\t{\tau}
\def\th{\theta}
\def\Th{\Theta}

\def\vep{\varepsilon}
\def\vp{\varphi}
\def\pa{\partial}
\def\un{\underline}
\def\ov{\overline}
\def\fr{\frac}
\def\sq{\sqrt}
\def\ot{\otimes}
\def\tf{\textbf}
\def\WW{\begin{stack}{\circle \\ W}\end{stack}}
\def\ww{\begin{stack}{\circle \\ w}\end{stack}}
\def\st{\stackrel}
\def\Ra{\Rightarrow}
\def\R{{\mathbb R}}
\def\mf{\mathbf }
\def\bi{\begin{itemize}}
\def\ei{\end{itemize}}
\def\i{\item}
\def\bt{\begin{tabular}}
\def\et{\end{tabular}}
\def\lf{\leftarrow}
\def\nn{\nonumber}
\def\va{\vartheta}
\def\wh{\widehat}
\def\vs{\vspace}
\def\Lam{\Lambda}
\def\sm{\setminus}
\def\ba{\begin{array}}
\def\ea{\end{array}}
\def\bd{\begin{description}}
\def\ed{\end{description}}
\def\lan{\langle}
\def\ran{\rangle}
\def\l{\label}
\def\mb{\mathbb}
\def\ti{\times}
\def\mc{\mathcal}
\def\v{\vec}
\large

\preprint{}
\title{ Measurement-Induced Non locality in an $n$-partite quantum state.}

\date{\today}
\author{Ali Saif M. Hassan}
\email{alisaif73@gmail.com}
 \affiliation{Department of Physics, University of Amran, Amran, Yemen}

\author{Pramod S. Joag}
\email{pramod@physics.unipune.ac.in}
\affiliation{Department of Physics, University of Pune, Pune, India-411007.}

\date{\today}
\begin{abstract}

We generalize the concept of measurement-induced non-locality (MiN) to $n$-partite quantum states. We get exact analytical expressions for MiN in an $n$-partite pure and $n$-qubit mixed state. We obtain the conditions under which MiN equals geometric quantum discord in an $n$-partite pure state and an $n$-qubit mixed state.

\noi PACS numbers: 03.65.Ud;75.10.Pq;05.30.-d
 \end{abstract}

\maketitle

Measurement induced non-locality (MiN) is a measure of quantum correlations as manifested in the non-local effects of local (on a single part) quantum operations\cite{fu,luo}. These local quantum operations leave invariant the reduced density operators of the parts on which they act, while changing the global quantum state. MiN concerns the von-Neumann measurement on a part of a quantum system.
MiN being an inherently quantum phenomenon, is expected to be useful as a tool for quantitative  specification of quantum correlation. Such a quantitative specification of quantum correlations in terms of MiN was given in \cite{luo} for bipartite quantum systems. Here we generalize this measure to $n$-partite quantum systems. MiN is a manifestation of the quantum verses classical paradigm of quantum correlations and naturally compares with quantum discord \cite{olli,hend,dakic,joag} which is also a manifestation of such a paradigm.  In fact, it is quite relevant to inquire about the conditions on quantum states under which MiN and geometric discord are equal (or, rather are different) and the different kinds of information they give about the quantum correlations in a quantum state. Here we establish such general conditions in $n$-partite pure and $n$-qubit mixed states.\\

To understand the non-local effects involved, consider a bipartite quantum system. The state of a bipartite system may be changed by locally invariant operation applied to one of the subsystems. This change in the bipartite state is a non-local effect and can be detected only by measuring the two parts jointly. By employing a Hilbert-Schmidt metric, for example, we can quantify such non-local effects by measuring the distance between initial and final bipartite states. These ideas are further clarified by considering an application like quantum dense coding. In this process two parties share an entangled pair of qubits (in the Bell state) one of which is subjected to a local unitary operation which does not change its reduced density operator. In other words, the marginal statistics of measurements on the particle does not change by the local operation applied to it. Thus the reduced density operators of both the qubits do not change in the process. However, the state of the whole system (the bipartite state) changes after the local unitary operation is applied to one of the qubits. Thus the change in the state of the whole system due to a local operation on a part is a non-local effect and and can be observed only by measuring the two qubits jointly. There is no way to detect this change locally, that is, there is no way for any eavesdropper to succeed by dealing with only one of the two qubits. Further, this is essentially quantum non-locality as it necessarilly involves a pair of entangled qubits in a bipartite pure state. The relation of such a nonlocality with other measures of quantum correlations is a naturally interesting question. In this paper, we address this question by exploring the relation of MiN with discord and entanglement in an $n$-partite quantum system.\\

We may note here that the processes defining discord and MiN naturally devide the $n$-partite system into two parts, one subjected to measurement and the remaining part. However, MiN measures the change in the $n$-partite state brought out by such a local measurement (see Eq.(\r{e1})) and is related to the multipartite correlations implied by it (see theorems 1,2,3 below). The same statement applies to discord as well \cite{joag}. Thus both MiN and discord are amenable to genuine multipartite generalization.\\   

\emph{Multipartite generalization of MiN} :
 Multipartite generalization of MiN can be obtained in a manner analogous to that of geometric quantum discord \cite{joag}. For an $n$-partite system in a state $\rho$ we define, for (normalized) MiN \cite{giro}
 \ben   \l{e1}
N_{l}(\rho)=\fr{d_l}{d_l-1}\max_{\Pi^{(l)}} (||\rho-\Pi^{(l)}(\rho)||^2),l=1,2,\cdots,n
\een
where $\Pi^{(l)}=\{\Pi_k^{(l)}\}$ stands for the set of von-Neumann measurements on the $l$th part such that $\Pi^{(l)}(\rho^{(l)}) = \sum_{k} \Pi_k^{(l)}\rho^{(l)} \Pi_k^{(l)} = \rho^{(l)}$, $\rho^{(l)}$ being the reduced density operator obtained by tracing out all parts other than the $l$th part from the $n$-partite state acting on $\mathcal{H}=\mathcal{H}^{1}\otimes \mathcal{H}^{2} \otimes \cdots \otimes \mathcal{H}^{n}$ with $dim(\mathcal{H}^{m}) = d_m, \; m=1,2,\cdots,n$. Such a measurement $\Pi^{(l)}$ is defined by the projectors corresponding to the eigenstates of $\rho^{(l)}$. When all the eigenvalues of $\rho^{(l)}$ are non-degenerate, there is only one von-Neumann measurement $\Pi^{(l)}$ satisfying $\Pi^{(l)}(\rho^{(l)}) = \sum_{k} \Pi_k^{(l)}\rho^{(l)} \Pi_k^{(l)} = \rho^{(l)}$ and the maximization requirement in Eq.(\r{e1}) drops out. If one or more eigenvalues of $\rho^{(l)}$ are degenerate, the right hand side of Eq.(\r{e1}) has to be maximized over the eigenspaces of degenerate eigenvalues, which is, in general, a difficult task.

Throughout this paper the superscript $t$ denotes the transpose of a vector or a matrix.

Comparing the definitions of MiN $N_{l}(\rho)$ and the geometric discord $D_{l}(\rho)$ \cite{joag} it follows that, for any $n$-partite state, $ N_{l}(\rho)\geq D_{l}(\rho).$ We are interested in finding the criteria for their equality.

The multipartite non-locality can be evaluated for an $n$-partite pure state via the following

\emph{Theorem 1}: Let $|\psi\ran=\sum_{i_1i_2\cdots i_n} a_{i_1i_2\cdots i_n}|i_1i_2\cdots i_n\ran$ be a $n$-partite pure state. Then
 \ben   \l{e2}
N_{l}(|\psi\ran \lan \psi|)=\fr{d_l}{d_l-1}(1-tr(\rho^{(l)})^2),
\een
where $\rho^{(l)}$ is the reduced density matrix of the $l$th part and $d_l=dim(H^{l}).$ 

\emph{Proof}: In order to get $N_{l}(|\psi\ran \lan \psi|)$ we can directly calculate the terms which define it (Eq.(\r{e1})). We have
\ben \l{eq5}
\rho=|\psi\ran \lan \psi|=\sum_{i_1i_2\cdots i_n}\sum_{j_1j_2\cdots j_n} a_{i_1i_2\cdots i_n} a^*_{j_1j_2\cdots j_n}|i_1i_2\cdots i_n\ran \lan j_1j_2\cdots j_n|.\\
\een
Here $|i_1i_2\cdots i_n\ran$ is the orthonormal product basis in the $n$-partite Hilbert space. The set of von-Neumann measurements on the $l$th part is given by $$\Pi^{(l)}=\{\Pi_k^{(l)}=U|k_l\ran \lan k_l|U^{\dagger} \}$$
where $\{|k_{l}\ran\},\;k_{l}=1,\ldots ,d_l=dim(H^{(l)})$ is an orthonormal basis in $H^{(l)}$ and $U$ is a unitary operator acting on $H^{(l)}.$ We can span all orthonormal bases in $H^{(l)}$ by varying $U.$ The post measurement state (after measurement on the $l$th part) is
\ben \l{eq3}
\Pi^{(l)}(\rho)=\sum_{k_l}^{d_l}\Pi^{(l)}_{k_l}(\rho)\Pi^{(l)}_{k_l}  \\
\een
where $\Pi^{(l)}_{k_l} = I_{d_1}\ot I_{d_2}\ot \cdots \ot I_{d_{l-1}}\ot \Pi_{k_l} \ot I_{d_{l+1}}\ot \cdots \ot I_{d_n}.$ We need $tr(\rho\Pi^{(l)}(\rho)).$   
A direct calculation of $tr(\rho\Pi^{(l)}(\rho))$ and comparison with $\rho^{(l)}=tr_{\bar{l}}(\rho)$ gives, assuming that $\{U|k_{l}\ran\}$ is the eigenbasis of $\rho^{(l)},$
\ben \l{eq4}
tr(\rho\Pi^{(l)}(\rho))=\sum_{k_l} (\lan k_l|U^{\dagger}\rho^{(l)}U|k_l\ran)^2=\sum_{k_l} \lambda_{k_l}^2=tr(\rho^{(l)})^2,   \\
\een
 where $\{\lambda_{k_l}\}$ are the eigenvalues of $\rho^{(l)}$. This calculation is done in the appendix.

From the definition of $N_l(\rho)$ (Eq.(\r{e1})) we get
$$ N_{l}(\rho)=\fr{d_l}{d_l-1}(||\rho||^2-\min_{\Pi^{(l)}}(2 tr(\rho \Pi^{(l)}(\rho))-||\Pi^{(l)}(\rho)||^2))).$$
For a pure state $||\rho||^2=1$ and $||\Pi^{(l)}(\rho)||^2=tr(\rho \Pi^{(l)}(\rho))$ so that
$$N_{l}(\rho)=\fr{d_l}{d_l-1}(1-\min_{\Pi^{(l)}}tr(\rho \Pi^{(l)}(\rho))).$$

The minimum is over the von-Neumann measurements leaving the marginal state $\rho^{(l)}$ invariant, that is $\sum_{k} \Pi_k^{(l)}\rho^{(l)} \Pi_k^{(l)} = \rho^{(l)}$, or,
$$\sum_{k_l}\lan k_l|U^{\dagger} \rho^{(l)}U|k_l\ran U|k_l\ran \lan k_l|U^{\dagger} =\rho^{(l)}.$$
This is the spectral decomposition of $\rho^{(l)}$ which is consistent with our choice of $\{U|k_l\ran\}$ to be the eigenbasis of $\rho^{(l)}$. Since $tr(\rho \Pi^{(l)}(\rho))$ is simply the trace of $(\rho^{(l)})^2$, the minimization in the definition of $N_l$ Eq.(\r{e1}) drops out and we get
$$N_{l}(|\psi\ran \lan \psi|)=\fr{d_l}{d_l-1}(1-tr(\rho^{(l)})^2).$$

\emph{Corollary}: For an $n$-partite pure state $\rho=|\psi\ran \lan \psi|$
\ben   \l{e3}
D_l(\rho)=N_l(\rho)
\een
where $D_l(\rho)$ is the geometric discord of $\rho$ with von-Neumann measurement on the $l$th part \cite{joag}.
This important result follows trivially, because $D_l(\rho)$ requires maximization over all von-Neumann measurements on the $l$th part which is obtained if the $\{U|k_l\ran\}$ forms the eignbasis of $\rho^{(l)}$. To make it more explicit, note that, for a pure state, 
$$D_l(\rho)=\frac{d_{l}}{2(d_{l}-1)}(1-\max_{\Pi^{(l)}}tr(\rho \Pi^{(l)}(\rho)))=\frac{d_{l}}{2(d_{l}-1)}(1-\max_{U|k_{l}\ran}tr(\rho \Pi^{(l)}(\rho))),$$
where the maximization is over all von Neuman measurements on the $l$th part. We get the the maximization in the second term only when $\{U|k_{l}\ran \}$ form the eigenbasis of $\rho^{(l)},$ so $$D_{l}(\rho)=\fr{d_l}{d_l-1}(1-\sum_{k_{l}}\la^{2}_{k_{l}}) =\fr{d_l}{d_l-1}(1-tr(\rho^{(l)})^2)=N_{l}(\rho).$$
 
It is interesting to compare $N_{l}(\rho)$ (Eq.(\r{e1})) with measures of entanglement of pure multipartite states. For a bipartite pure state $\rho_{AB}$ we have, for the concurrence, $$C(\rho_{AB})=\sqrt{2(1-tr(\rho_{A}^2))}$$ which is related to $N_{l}(\rho_{AB})$ by 
\ben \l{eq1}
N_{l}(\rho_{AB})=\frac{d_{l}}{2(d_{l}-1)} C^2(\rho_{AB}).
\een
Thus, for pure bipartite states, non-locality is simply related to concurrence.

The Meyer-Wallach measure of entanglement of multipartite pure states is $$Q(|\psi\ran)=\frac{1}{n}\sum_{k=1}^{n}2(1-tr(\rho_{k}^2))$$ where $\rho_{k}$ is the reduced density operator for the $k$th part. Thus, 
\ben \l{eq2}
Q(|\psi\ran)=\frac{2}{n}\sum_{l=1}^n\left(\frac{d_{l}-1}{d_{l}}\right) N_{l}(|\psi\ran \lan \psi|).
\een
Thus the Meyer-Wallach measure of pure state multipartite entanglement is the average of non-locality over the parts of the system.

\emph{Non-locality in the multipartite mixed states}:
To get $N_l(\rho)$ in this case, we start with the Bloch representation of a multipartite state $\rho$\cite{ali1}.
Bloch representation \cite{ali1} of a $n$-partite density operator is

\ben \l{e4}
\rho=\fr{1}{\Pi_k^n d_k} \{\otimes_k^n I_{d_k}+ \sum_{k \in \mathcal{N}}\sum_{\alpha_{k}}s_{\alpha_{k}}\lambda^{(k)}_{\alpha_{k}}
+\sum_{2\leq M \leq n}\sum_{\{k_1,k_2,\cdots,k_M\}}\sum_{\alpha_{k_1}\alpha_{k_2}\cdots \alpha_{k_M}}\tilde{t}_{\alpha_{k_1}\alpha_{k_2}\cdots \alpha_{k_M}}\lambda^{(k_1)}_{\alpha_{k_1}} \lambda^{(k_2)}_{\alpha_{k_2}}\cdots \lambda^{(k_M)}_{\alpha_{k_M}}\}
\een
 where $\mathcal{N}=\{1,2,\cdots,n\}$ and

 $$ \lambda^{(k_1)}_{\alpha_{k_1}}=(I_{d_1}\otimes I_{d_2}\otimes \dots \otimes \lambda_{\alpha_{k_1}}\otimes I_{d_{k_1+1}}\otimes \dots \otimes I_{d_n})$$
 $$\lambda^{(k_2)}_{\alpha_{k_2}}=(I_{d_1}\otimes I_{d_2}\otimes \dots \otimes \lambda_{\alpha_{k_2}}\otimes I_{d_{k_2+1}}\otimes \dots \otimes I_{d_n})$$
\ben \l{e5}
  \lambda^{(k_1)}_{\alpha_{k_1}} \lambda^{(k_2)}_{\alpha_{k_2}}=(I_{d_1}\otimes I_{d_2}\otimes \dots \otimes \lambda_{\alpha_{k_1}}\otimes I_{d_{k_1+1}}\otimes \dots \otimes \lambda_{\alpha_{k_2}}\otimes I_{d_{k_2+1}}\otimes I_{d_n})
\een
   $\textbf{s}^{(k)}$ is a Bloch vector corresponding to $k$th subsystem,  $\textbf{s}^{(k)} =[s_{\alpha_{k}}]_{\alpha_{k}=1}^{d_k^2-1} $ and
\ben \l{e6}
   \tilde{t}_{\alpha_{k_1}\alpha_{k_2}\dots\alpha_{k_M}}=\fr{d_{k_1}d_{k_2}\dots d_{k_M}}{2^M} tr[\rho \lambda^{(k_1)}_{\alpha_{k_1}} \lambda^{(k_2)}_{\alpha_{k_2}}\cdots \lambda^{(k_M)}_{\alpha_{k_M}}].
   \een
For more details see ref.\cite{ali1,mahl,kim}.

Recently, for a bipartite system $ab$ $(n=2)$ with states in $\mc{H}^a \ot \mc{H}^b ,\;dim(\mc{H}^a)=d_a,\;dim(\mc{H}^b)=d_b ,$ S. Luo and S. Fu introduced the following generic expression for MiN \cite{luo}
\ben  \l{e7}
N_a (\rho) = tr(TT^t )- \min_{A} tr(ATT^tA^t ),
\een
where $T = [t_{ij} ]$ is an $d_a^2\ti d_b^2$ matrix and the minimum is taken over all $(d_a\times d_a^2-1)$-dimensional isometric matrices $A = [a_{ji} ]$ such that $a_{ji}  = tr(|j\rangle\langle j|X_i)=\lan j|X_i|j\ran,\;\; j = 1,2,\ldots,d_a\;;\{X_i \},i = 1,2,\ldots,d_a^2-1$ forms an orthonormal basis in the space of operators acting on $\mc{H}^a$ and $\{|j\rangle\}$ is any orthonormal basis in $\mc{H}^a$. we generalize this result to $n$-partite quantum states, in theorem 2 and 3.

\emph{Theorem 2}. Let $\rho_{12\cdots n}$ be a $n$-partite state defined by Eq.(\r{e4}), then
\ben  \l{e8}
N_l(\rho)=\fr{d_l}{(d_l-1)\Pi_k^n d_k}\{\sum_{1\leq M \leq n-1}\sum_{\{k_1,k_2,\cdots,k_M\} \subseteq \mathcal{N}-\{l\}} \fr{d_ld_{k_1}d_{k_2}\cdots d_{k_M}}{2^{M+1}} ||\mathcal{T}^{\{l,k_1,k_2,\cdots,k_M\}}||^2-\min_{A^{(l)}} tr(A^{(l)} K^{(l)} (A^{(l)})^t)\},
\een
where the $(d_l^2-1)\times (d_l^2-1)$ symmetric matrix $K^{(l)}$ is defined as
$$ K_{\alpha_l\beta_l}^{(l)}=\sum_{1\leq M \leq n-1}\sum_{\{k_1,k_2,\cdots,k_M\} \subseteq \mathcal{N}-\{l\}} \sum_{\alpha_{k_1}\alpha_{k_2}\cdots \alpha_{k_M}} \fr{d_ld_{k_1}d_{k_2}\cdots d_{k_M}}{2^{M+1}} t_{\alpha_l\alpha_{k_1}\alpha_{k_2}\cdots \alpha_{k_M}}  t_{\beta_l\alpha_{k_1}\alpha_{k_2}\cdots \alpha_{k_M}}.$$
 $\mathcal{T}=[t_{i_1 i_2 \cdots i_M}]=[tr(\rho \lambda^{(k_1)}_{\alpha_{k_1}} \lambda^{(k_2)}_{\alpha_{k_2}}\cdots \lambda^{(k_M)}_{\alpha_{k_M}})]$ and the maximum is taken over all $d_{l}\times (d_{l}^2-1)$ dimensional matrices $A^{(l)}=[a_{ji_l}],$ such that $a_{ji_l}=tr(|j\rangle\langle j| \fr{\lambda^{(l)}_{i_l}}{\sqrt2}), \;j=1,2,\ldots,d_{l};\;i_l=1,2,\ldots,d_{l}^2-1$ and $\{|j\rangle\}$ is any orthonormal basis for $\mathcal{H}^{(l)}.$ In particular, we have
\ben  \l{e9}
N_l(\rho)\leq \sum_{i=1}^{d_l^2-d_l} \eta_i,
\een
where$\{\eta_i:i=1,2,\cdots, d_l^2-1\}$ are the eigenvalues of the $(d_l^2-1)\times (d_l^2-1)$ symmetric matrix $K^{(l)}$ listed in non-increasing order. Furthermore, if $\rho^{(l)}=tr_{\bar{l}}\rho_{12\cdots n}$ is non-degenerate with spectral projections $\{|j\rangle\langle j|\}$, then
\ben  \l{e10}
N_l(\rho)=\fr{d_l}{(d_l-1)\Pi_k^n d_k}\{\sum_{1\leq M \leq n-1}\sum_{\{k_1,k_2,\cdots,k_M\} \subseteq \mathcal{N}-\{l\}} \fr{d_ld_{k_1}d_{k_2}\cdots d_{k_M}}{2^{M+1}} ||\mathcal{T}^{\{l,k_1,k_2,\cdots,k_M\}}||^2-tr(A^{(l)} K^{(l)} (A^{(l)})^t)\}.
\een

%The proof of this theorem is a straight-forward generalization of theorem (2) in ref. [] to the multipartite case, so that we skip it.

\emph{Theorem 3}. If the $l$th part of a $n$-partite quantum system is a qubit $(d_l=2)$, then
\begin{displaymath}
N_{l}(\rho)=\fr{d_l}{(d_l-1)\Pi_k^n d_k}\left[\sum_{1 \leq M \leq n-1} \sum_{\{k_1,\ldots,k_M\} \subseteq \mathcal{N}-\{l\}} \fr{d_{k_1}d_{k_2}\cdots d_{k_M}}{2^{M}} ||\mathcal{T}^{\{l,k_1,\ldots,k_M\}}||^2- \left\{ \begin{array}{ll}
  \frac{{\textbf{s}^{(l)}}^t K^{(l)}\textbf{s}^{(l)}}{||\textbf{s}^{(l)}||^2},  & \textrm{if $\textbf{s}^{(l)}\neq 0$}\\
 \eta_{min},  & \textrm{if $\textbf{s}^{(l)}= 0$}
\end{array} \right. \right] \eqno{(11a)}
\end{displaymath}

where $\textbf{s}^{(l)}$ is the coherent vector of $\rho^{(l)}$ and $\eta_{min}$ is the smallest eigenvalue of the matrix $K^{(l)}$ which is a $3 \times 3$ real symmetric matrix, defined as
\ben \l{e11}
K_{\al_{l}\b_{l}}^{(l)}= \sum_{1\leq M \leq n-1}\sum_{\{k_1,\ldots,k_M\} \subseteq \mathcal{N}-\{l\}}\sum_{\alpha_{k_1}\alpha_{k_2}\cdots \alpha_{k_M}}   \fr{d_{k_1}d_{k_2}\cdots d_{k_M}}{2^{M}} t_{\alpha_{l}\alpha_{k_1}\alpha_{k_2}\cdots \alpha_{k_M} } t_{ \beta_{l}\alpha_{k_1}\alpha_{k_2}\cdots \alpha_{k_M}} .
\een
For $n$-qubit $(d_i=2,\;i=1,2,\cdots,n)$,
\begin{displaymath}
N_{l}(\rho)=\frac{1}{2^{(n-1)}}\left[ \sum_{1 \leq M \leq n-1}\sum_{\{k_1,\ldots,k_M\} \subseteq \mathcal{N}-\{l\}}  ||\mathcal{T}^{\{l,k_1,\ldots,k_M\}}||^2- \left\{ \begin{array}{ll}
  \frac{{\textbf{s}^{(l)}}^t K^{(l)}\textbf{s}^{(l)}}{||\textbf{s}^{(l)}||^2},  & \textrm{if $\textbf{s}^{(l)}\neq 0$}\\
 \eta_{min},  & \textrm{if $\textbf{s}^{(l)}= 0$}
\end{array} \right.\right]
\end{displaymath}
and
\ben \l{e12}
K_{\al_{l}\b_{l}}^{(l)}= \sum_{1\leq M \leq n-1}\sum_{\{k_1,\ldots,k_M\} \subseteq \mathcal{N}-\{l\}}\sum_{\alpha_{k_1}\alpha_{k_2}\cdots \alpha_{k_M}} t_{\alpha_{l}\alpha_{k_1}\alpha_{k_2}\cdots \alpha_{k_M} } t_{ \beta_{l}\alpha_{k_1}\alpha_{k_2}\cdots \alpha_{k_M}} .
\een
 The proofs of theorems 2 and 3 is a straightforward generalization of those of theorems 2 and 3 respectively in ref. \cite{luo} to the multipartite case, so that we skip these proofs.

\emph{Relation between the non-locality and geometric quantum discord for arbitrary $n$-qubit states} : We saw (see Eq.(\r{e2})) that the non-locality and geometric discord are equal for arbitrary $n$-partite pure states. In this section we find a class of general $n$-qubit states for which these quantities coincide.
Consider a $n$-qubit state $\rho$. The geometric discord for such a state corresponding to the von-Neumann measurement on $l$th qubit is given by
\ben \l{e13}
D_{l}(\rho)=\frac{1}{2^{(n-1)}}\left[||s^{(l)}||^2+ \sum_{1 \leq M \leq n-1}\sum_{\{k_1,\ldots,k_M\} \subseteq \mathcal{N}-\{l\}}  ||\mathcal{T}^{\{l,k_1,\ldots,k_M\}}||^2- \lambda_{max},
\right]
\een
where $s^{(l)}$ is the coherent vector of $\rho^{(l)}$ (reduced density operator for the $l$th part), $\mathcal{T}= [t_{\alpha_{k_1}\alpha_{k_2}\dots\alpha_{k_M}}]=[tr(\rho \lambda^{(k_1)}_{\alpha_{k_1}} \lambda^{(k_2)}_{\alpha_{k_2}}\cdots \lambda^{(k_M)}_{\alpha_{k_M}})]$, and  $\lambda_{max}$ is the largest eigenvalue of the $3\times 3$ real symmetric matrix
\ben \l{e14}
 G^{(l)}= \textbf{s}^{(l)}(\textbf{s}^{(l)})^t+K^{(l)}
 \een
where $K^{(l)}$ is given by Eq.(\r{e12}) for $n$ qubits.
The non-locality for the $n$-qubit state $\rho$ is given by Eq.(11a). We now consider two cases

Case I : $\textbf{s}^{(l)}\neq 0$. By Eq.(\r{e14}) we get $$ \hat{e}^t G^{(l)} \hat{e}= \hat{e}^t \textbf{s}^{(l)}(\textbf{s}^{(l)})^{t}\hat{e}+\hat{e}^t K^{(l)}\hat{e}$$ where $\hat{e} \in R^3$ is an arbitrary unit vector, choosing $\hat{e}=\fr{\textbf{s}^{(l)}}{||\textbf{s}^{(l)}||},$ we get
$$\fr{(\textbf{s}^{(l)})^t K^{(l)}\textbf{s}^{(l)}}{||\textbf{s}^{(l)}||^2}=\fr{(\textbf{s}^{(l)})^t G^{(l)}\textbf{s}^{(l)}}{||\textbf{s}^{(l)}||^2}-||\textbf{s}^{(l)}||^2$$
Substituting in Eq.(11a) we get
\ben \l{e15}
N_{l}(\rho)=\frac{1}{2^{(n-1)}}\left[||\textbf{s}^{(l)}||^2 +\sum_{1 \leq M \leq n-1}\sum_{\{k_1,\ldots,k_M\} \subseteq \mathcal{N}-\{l\}}  ||\mathcal{T}^{\{l,k_1,\ldots,k_M\}}||^2- \frac{{\textbf{s}^{(l)}}^t G^{(l)}\textbf{s}^{(l)}}{||\textbf{s}^{(l)}||^2}\right]
\een
If $\fr{\textbf{s}^{(l)}}{||\textbf{s}^{(l)}||}$ is the eigenvector of $G^{(l)}$ with the largest eigenvalue then the right hand side of Eq.(\r{e14}) gives the geometric discord $D_l(\rho)$ so that under this condition $N_l(\rho)=D_l(\rho)$. The above condition can be equivalently stated as $$[\textbf{s}^{(l)}(\textbf{s}^{(l)})^t,K^{(l)}]=0$$ and $$||\textbf{s}^{(l)}||^2+\eta_l \geq \eta_{i\neq l}$$ where $\{\eta_i\}$ are the eigenvalues of $K^{(l)}$ and  $\eta_l$ is the eigenvalue corresponding to the eigenvector $\fr{\textbf{s}^{(l)}}{||\textbf{s}^{(l)}||}.$

Case II: $\textbf{s}^{(l)}=0$. In this case $\rho$ has one doubly degenerate eigenvalue. With  $\textbf{s}^{(l)}=0$ we get from Eq.(\r{e14})
\ben \l{e16}
\hat{e}^t G^{(l)} \hat{e}= \hat{e}^t K^{(l)}\hat{e}.
\een
 To get non-locality we have to minimize the right hand side while the geometric discord requires maximization of the left hand side. Under these conditions, the equality in Eq.(\r{e15}) is preserved if $G^{(l)}=K^{(l)}$ has a single three-fold degenerate eigenvalue,$(\eta_1=\eta_2=\eta_3).$ Thus when $\textbf{s}^{(l)}=0,$ $N_l(\rho)=D_l(\rho)$ provided the matrix $K^{(l)}$ has a single three fold degenerate eigenvalue.

\begin{figure}[!ht]
\begin{center}
\includegraphics[width=8cm,height=5cm]{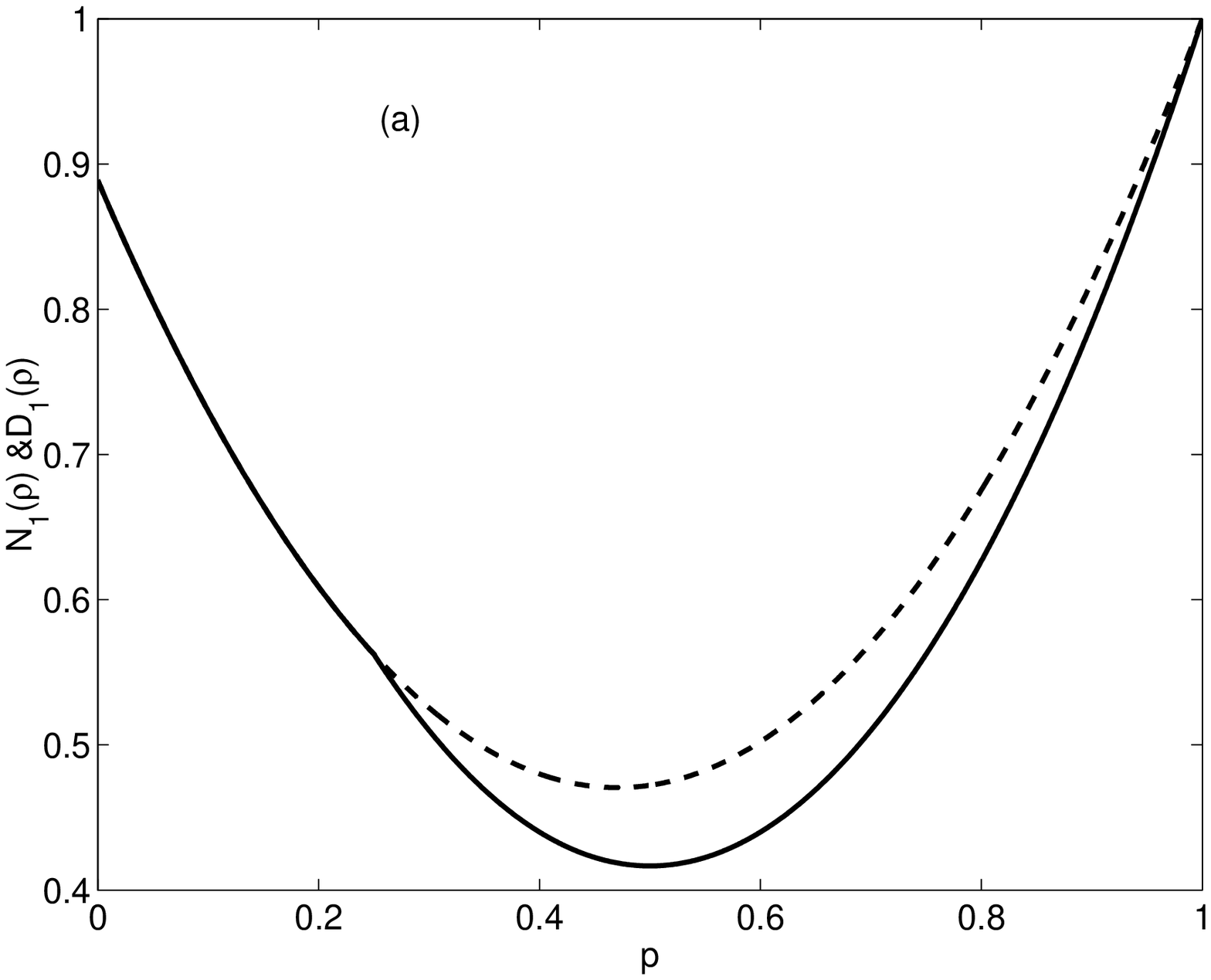}
\includegraphics[width=8cm,height=5cm]{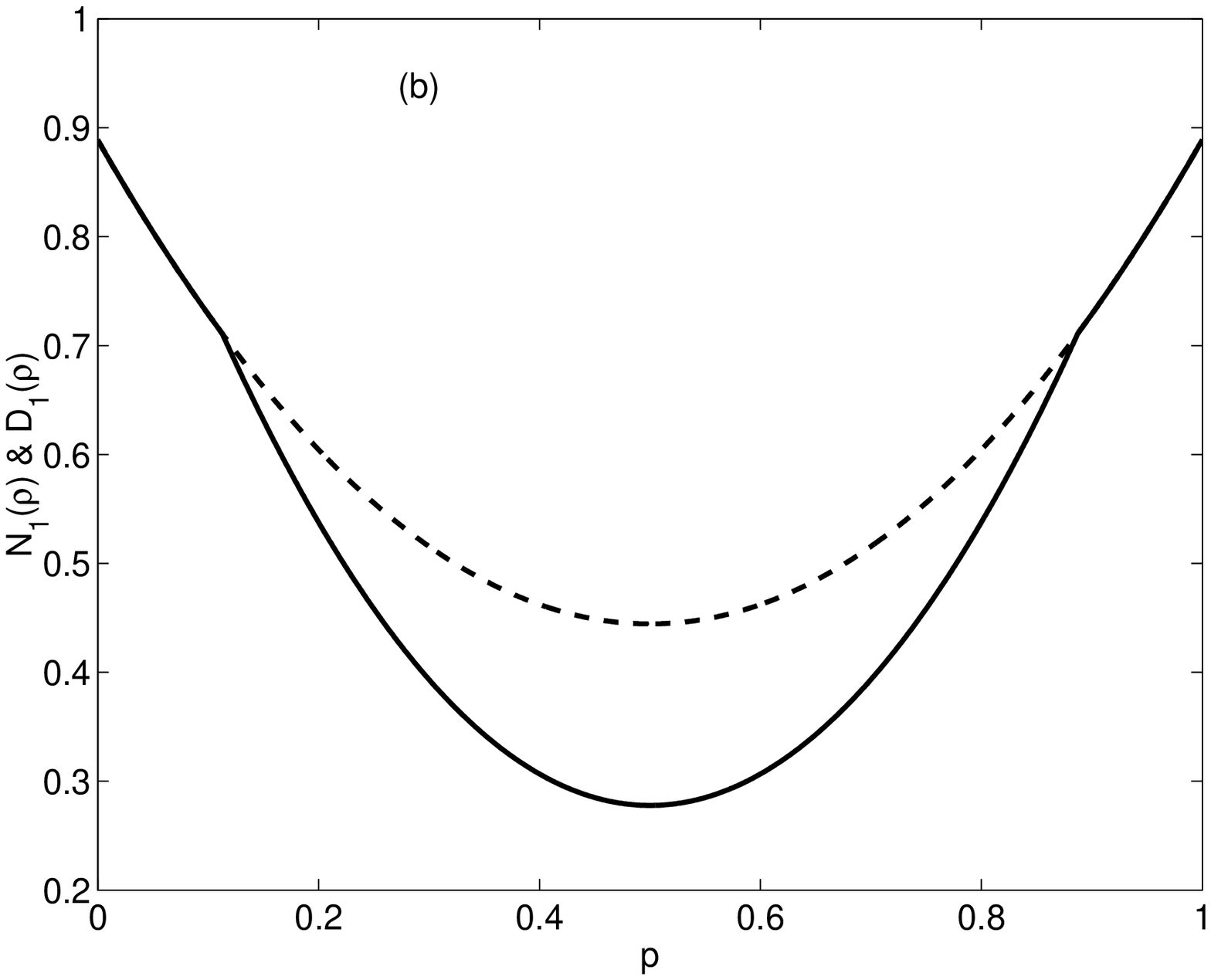}
\includegraphics[width=8cm,height=5cm]{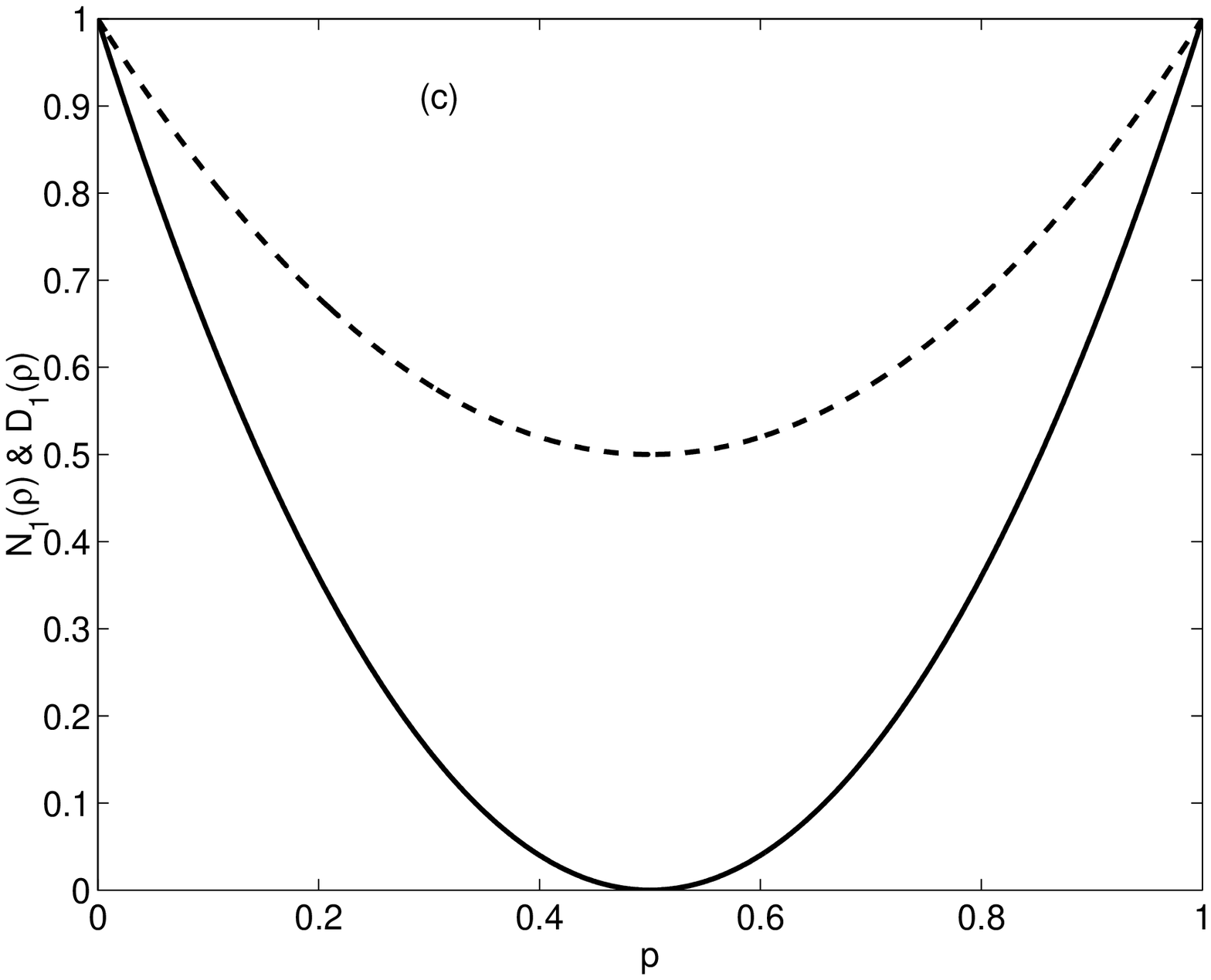}
\includegraphics[width=8cm,height=5cm]{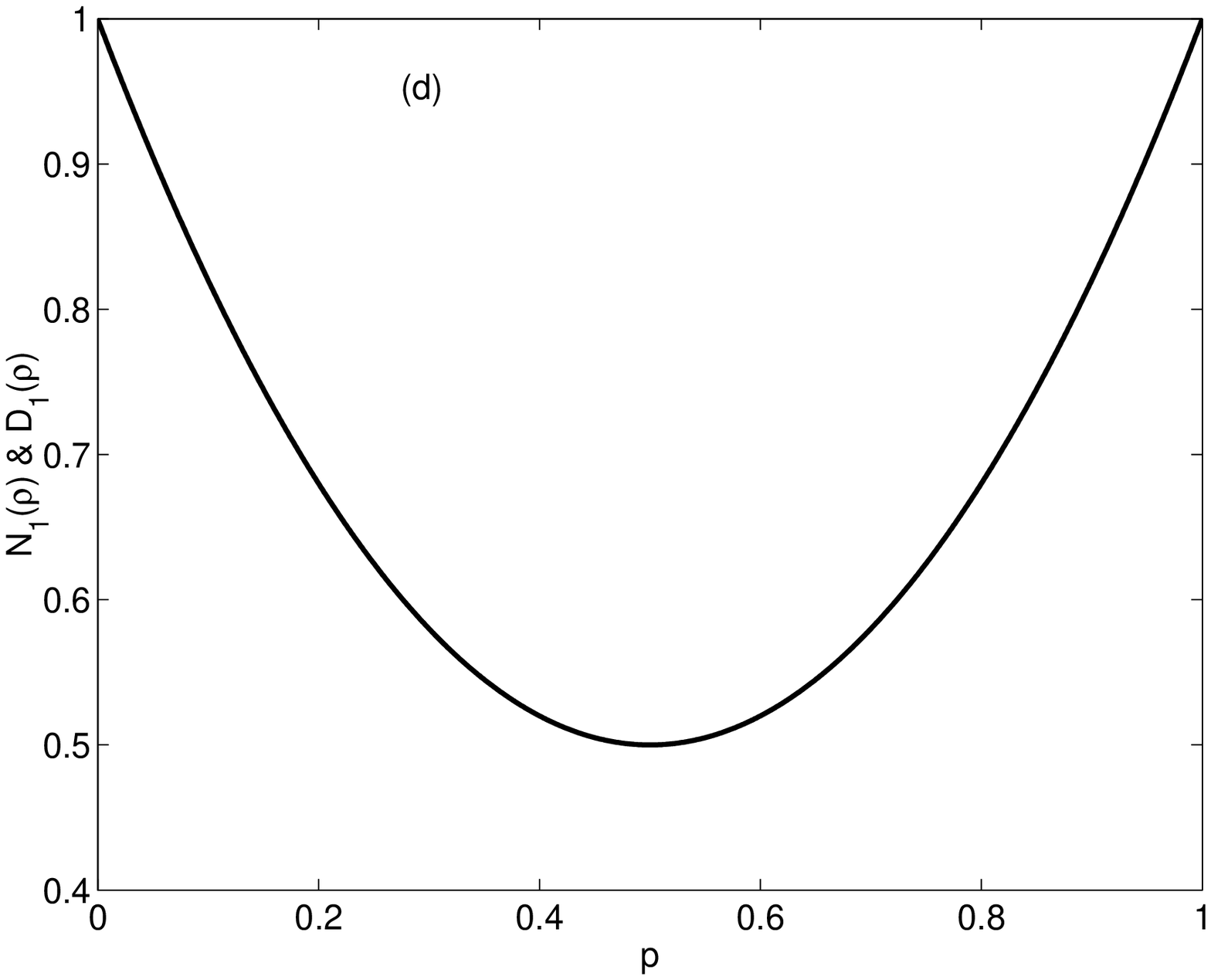}
Figure 1 : The variation of MiN $N_l(\rho(p))$ (dashed carve) and the geometric discord $D_l(\rho(p))$ (continuous carve) for the state as in (a)  Eq(\r{e17})  (b) Eq(\r{e18}) (c) Eq(\r{e19})   (d) Eq(\r{e20}) with parameter $p$.
\end{center}
\end{figure}

\emph{Examples} : As our first example we consider the set of three qubit states comprising the convex combination of $|GHZ\ran=\fr{1}{\sqrt{2}}(|000\ran +|111\ran)$ and $|W\ran=\fr{1}{\sqrt{3}}(|001\ran +|010\ran+|100\ran),$
\ben \l{e17}
\rho(p)=p|GHZ\ran \lan GHZ|+(1-p)|W\ran \lan W|.
\een
The $K^{(1)}$ matrix of this state is

 $$K^{(1)} = \textrm{diag}[ 2p^2+\fr{16}{9}(1-p)^2, 2p^2+\fr{16}{9}(1-p)^2,  2p^2+\fr{19}{9}(1-p)^2-\fr{4}{3}p(1-p)]$$
with the coherent vector for the first qubit  $$\textbf{s}^{(1)} =[  0, 0,\fr{1}{3}(1-p)]^t\neq 0 $$
so that case I applies. We find that $[\textbf{s}^{(1)}(\textbf{s}^{(1)})^t,K^{(1)}]=0$  and the condition $||\textbf{s}^{(1)}||^2+\eta_1 \geq \eta_{i\neq 1},$ ($\eta_1$ is the eigenvalue of $K^{(1)}$ matrix corresponding to eigenvector $\fr{ \textbf{s}^{(1)}}{||\textbf{s}^{(1)}||}$), is satisfied when $p \leq \fr{1}{4}$ and $p=1$. This is depicted in fig.(1a).

 The second example consists of
 \ben \l{e18}
 \rho(p)=p|\widetilde{W}\ran \lan \widetilde{W}|+(1-p)|W\ran \lan W|
 \een
 where $|\widetilde{W}\ran$ is the flipped $|W\ran $ state, $\si_x\otimes \si_x \otimes \si_x |W\ran$. The $K^{(1)}$ matrix of this state is
 $$K^{(1)} = \textrm{diag}[\fr{16}{9}p^2+\fr{16}{9}(1-p)^2,  \fr{16}{9}p^2+\fr{16}{9}(1-p)^2,  \fr{19}{9}p^2+\fr{19}{9}(1-p)^2-\fr{10}{3}p(1-p)]$$
with the coherent vector for the first qubit  $$\textbf{s}^{(1)} = [ 0, 0, \fr{1}{3}(1-2p)]^t \neq 0,$$
so that case I applies. We find that $[\textbf{s}^{(1)}(\textbf{s}^{(1)})^t,K^{(1)}]=0$  and the condition $||\textbf{s}^{(1)}||^2+\eta_1 \geq \eta_{i\neq 1},$ is satisfied when $p \leq 0.1127$ and $p \geq 0.8873$. The results are shown in fig.(1b).

The third example consists of
\ben \l{e19}
\rho(p)=p|GHZ\ran \lan GHZ|+(1-p)|GHZ_-\ran \lan GHZ_-|
\een
 where $|GHZ_-\ran=\fr{1}{\sqrt{2}}(|000\ran-|111\ran).$ The $K^{(1)}$ matrix of this state is
 $$K^{(1)} = \textrm{diag}[2(2p^2-1)^2, 2(2p^2-1)^2, 2(2p^2+1)^2] $$
and the coherent vector for the first qubit  $$\textbf{s}^{(1)} = 0,$$
so that case II applies. $K^{(1)}$ does not have a single triply degenerate eigenvalue, for all $p$, except $p=0$ and $p=1$. Therefore $N_l(\rho) \neq D_l(\rho)$ for all $p$ between $0$ and $1$. The results are shown in fig.(1c).

 The last example consists of the states
 \ben \l{e20}
\rho(p)=p|GHZ\ran \lan GHZ|+(1-p)|GHZ_1\ran \lan GHZ_1|
\een
 where $|GHZ_1\ran=\fr{1}{\sqrt{2}}(|001\ran+|110\ran.$ The $K^{(1)}$ matrix of this state is
 $$K^{(1)} =  \textrm{diag}[2(p^2-(1-p)^2), 2(p^2-(1-p)^2), 2(p^2-(1-p)^2)]$$
and the coherent vector for the first qubit  $$\textbf{s}^{(1)} = 0,$$
so that case II applies. $K^{(1)}$ does have a single triply degenerate eigenvalue, for all $p$. Therefore $N_l(\rho) = D_l(\rho)$ for all $p$ as shown in fig.(1d).

\emph{Summary and comments} : In this paper, we have given exact analytical expressions for MiN in an $n$-partite pure and $n$-qubit mixed state. Apart from this we obtain two results which we think are useful in further understanding of quantum correlations in multipartite quantum systems. First we have shown that the geometric discord and MiN are equal for a multipartite pure quantum state. This indicates that, in the classical verses quantum scenario, quantum correlations and non-locality imply each other. This supprts the well known result that for a bipartite pure state, entanglement and non-locality are equivalent in the sense that an entangled bipartite pure state breaks bell inequality and vice versa. Further, we have shown that, for a bipartite pure state, concurrence can be obtained from MiN (Eq.(\r{eq1})), which, in turn, equals geometric discord. Thus for the bipartite pure states geometric discord and MiN do not give any new information on quantum correlations as compared to entanglement and quantum correlations seem to be essentially dominated by entanglement. Interestingly, for multipartite pure states, the Meyer-Wallach measure of entanglement is just the average over MiN with measurement on the $l$th part (Eq.(\r{eq2})). This equation points to a new relation between the entanglement and non-locality in a multipartite pure state and projects entanglement as a kind of average non-local effect. To the best of our knowledge, a quantitative relation between a measure of multipartite entanglement and a measure of non-locality has not appeared in the literature before.

In the light of our result relating MiN, discord and entanglement for the multipartite pure states (Eq.s(\r{e3},\r{eq1})), we can see that quantum non-locality and hence discord is complementary to entanglement for a two qubit pure state by comparing the teleportation and dense coding protocols. In the dense coding protocol, quantum non-locality is operating, because a local unitary operation encodes global information in the two qubit joint state which can be deciphered only by the joint measurement of the two entangled qubits. In the teleportation process \emph{we need to communicate two classical bits} to translate the state of the particle from one party to another far away party provided two parties are sharing maximally entangled state, while the dense coding process is the reverse of teleportation process, as we transfer one particle \emph{so as to communicate two classical bits} provided the parties are sharing maximally entangled state.

Whereas MiN and geometric discord are equal for \emph{all} multipartite pure states, these coincide only for a class of multiqubit mixed states. Thus, in general, discord and MiN show different characters in multiqubit mixed states. We have obtained exact analytical conditions necessary for the equality of MiN and quantum discord in a mixed multiqubit state. These conditions obtained in case I and case II above identify a class of states for which MiN and the geometric quantum discord coincide. Recently quantum discord is shown to measure the quantumness of the state rather than genuine quantum correlation \cite{Gess}. Thus for the class of states with equal $N_l(\rho)$ and $D_l(\rho),$ MiN seems to be identical or simply related to quantumness of the state. An understanding of relation between quantumness and non-locality by other routes will then be interesting. At any rate, it is interesting to explore the relation between $N_l(\rho)$ and $D_l(\rho)$ in the states for which they do not coincide, because these will improve our understanding of quantum correlations. In such situations MiN is distinct from quantumness and may even be independent of it \cite{mor}. Finaly, the results of this paper may be useful for a unified classification of correlations in a multipartite quantum state \cite{modi}. \\

\emph{Acknowledgments} : This work was supported by the BCUD grant RG-13. ASMH acknowledges University of Pune for hospitality during his visit when this work was carried out.\\

\emph{Appendix}:

We obtain Eq.(\r{eq4}). We use various symbols defined in the proof of theorem 1. Using Eq.s(\r{eq5},\r{eq3}) and the orthonormality of the product basis $|i_1i_2\cdots i_n\ran ,$ a bit lengthy but straightforward calculation gives
%\benr  \nonumber
\begin{multline}
tr(\rho\Pi^{(l)}(\rho)) = \sum_{k_l}\left[\sum_{q_l}\sum_{i_1\cdots i_l \cdots i_n}a_{i_1\cdots i_l \cdots i_n}a^{*}_{i_1\cdots q_l \cdots i_n}\lan k_l|U^{\dagger}|i_l\ran \lan|q_l|U|k_l\ran \right]       \\ 
 \left[\sum_{p_l}\sum_{j_1\cdots j_l \cdots j_n}a_{j_1\cdots p_l \cdots j_n}a^{*}_{j_1\cdots j_l \cdots j_n}\lan k_l|U^{\dagger}|p_l\ran \lan|j_l|U|k_l\ran \right]. \;\;\;\;\;\;\;\;(A1) 
\end{multline}
%\eenr
Now, we get for $\rho^{(l)}$
$$\rho^{(l)}=\sum_{j_l}\sum_{i_1\cdots i_l \cdots i_n}a_{i_1\cdots i_l \cdots i_n}a^{*}_{i_1\cdots j_l \cdots i_n}|i_l\ran\lan j_l|.  \eqno{(A2)}$$
If $\{|\psi_{q}\ran \}$ are the eigenvectors of $\rho^{(l)},$ then by spectral theorem we can write
$$\rho^{(l)}=\sum_{q}\lan\psi_{q}|\rho^{(l)}|\psi_{q}\ran |\psi_{q}\ran \lan\psi_{q}|.  \eqno{(A3)}$$
We now put Eq.(A2) in Eq.(A3) and find $(\lan\psi_{q}|\rho^{(l)}|\psi_{q}\ran)^2 ,$ take $|\psi_{k_l}\ran = U|k_l\ran$ and compare with Eq.(A1) to get 
$$tr(\rho\Pi^{(l)}(\rho))= \sum_{k_l}(\lan k_{l}|U^{\dagger}\rho^{(l)}U|k_{l}\ran)^2 = \sum_{k_l}\lambda_{k_l}^2 = tr(\rho^{(l)})^2 . \eqno{(A4)}$$


\begin{thebibliography}{99}
\bibitem{fu}
L. B. Fu, Europhys. Lett. \textbf{75}, 1 (2006).
\bibitem{luo}
 S. Luo and S. Fu,  Phys. Rev. Lett. \tf{106}, 120401 (2012).
 \bibitem{olli}
 H. Ollivier and W. H. Zurek, Phys. Rev. Lett. \tf{88}, 017901 (2001).
\bibitem{hend}
 L. Henderson and V. Vedral, J. Phy. A \tf{34}, 6899 (2001).
\bibitem{dakic}
 B. Dakic, V. Vedral, and C. Brukner, Phys. Rev. Lett. \tf{105},190502 (2010).
\bibitem{joag}
Ali Saif M. Hassan and Pramod S. Joag, J. Phys. A: Math. Theor. \textbf{45}, 345301 (2012).
\bibitem{giro}
D. Girolami and G. Adesso, Phys. Rev. A 84, 052110 (2011).
\bibitem{ali1}
 Ali Saif M. Hassan and Pramod S. Joag, Quantum Inf. Comput. \textbf{8}, 773 (2008).
\bibitem{mahl}
G. Mahler, Volker A. Weberruss, \emph{Quantum networks} (Springer-Verlag Berlin Heidelberg 1995).
\textbf{8}, 773 (2008).
\bibitem{kim}
M. S. Byrd and n. Khaneja, Phys. Rev. A \textbf{68}, 062322 (2003); G. Kimura, Phys. Lett. A \textbf{314}, 339 (2003).
\bibitem{kold06}
 T. G. Kolda, ``Multilinear operator for higher order decompositions'', Tech. Report SAnD2006-2081, Sandia national Laboratories, Albuquerque, new Mexico and Livermore, Colifornia Apr. 2006.
\bibitem{lmv00}
L. De Lathauwer, B. De Moor and J. Vandewalle, {\it SIAM J. Matrix Anal. A.} \textbf{21}, 1253 (2000).
\bibitem{Gess}
M. Gessner, Elsi-Mari Laine, Heinz-Peter Breuer, J. Piilo, Phys. Rev. A \textbf{85}, 052122 (2012).
\bibitem{mor}
T.Mor, Int. J. Quantum. Inf. \textbf{4}, 161 (2006). 
\bibitem{modi}
K.Modi,T.Paterek,W.Son,V.Vedral,M.Williamson, Phys.Rev.Lett. \textbf{104}, 080501 (2010).
\end{thebibliography}
\end{document}